\documentclass[submission, Proceedings]{SciPost}

\hypersetup{
    colorlinks,
    linkcolor={red!50!black},
    citecolor={blue!50!black},
    urlcolor={blue!80!black}
}
\usepackage{bm}

\def\la{\langle}
\def\ra{\rangle}
\def\spa#1.#2{\left\langle#1\,#2\right\rangle}
\def\spb#1.#2{\left[#1\,#2\right]}
\def\spba#1.#2.#3{\left[#1|#2|#3\right\rangle}

\newcommand{\mi}{\scalebox{0.94}[1.0]{-}}

\DeclareMathOperator{\tr}{\mathrm{tr}}

\def\notag{\nonumber}

\newcommand\bref[1]{(\ref{#1})}

\usepackage[bitstream-charter]{mathdesign}
\DeclareSymbolFont{usualmathcal}{OMS}{cmsy}{m}{n}
\DeclareSymbolFontAlphabet{\mathcal}{usualmathcal}

\begin{document}

\begin{center}{\Large \textbf{
		Proving the dimension-shift conjecture
}}\end{center}

\begin{center}
Ruth Britto\textsuperscript{1,2},
Guy R.~Jehu\textsuperscript{1$\star$} and
Andrea Orta\textsuperscript{1}

{\bf 1} School of Mathematics and Hamilton Mathematical 
Institute, Trinity College Dublin, Ireland\\
{\bf 2} Institut de Physique Th\'eorique,
		Universit\'e Paris Saclay, France\\
* \href{mailto:jehu@maths.tcd.ie}{jehu@maths.tcd.ie}
\end{center}

\begin{center}
\today
\end{center}


\definecolor{palegray}{gray}{0.95}
\begin{center}
\colorbox{palegray}{
  \begin{tabular}{rr}
  \begin{minipage}{0.1\textwidth}
    \includegraphics[width=35mm]{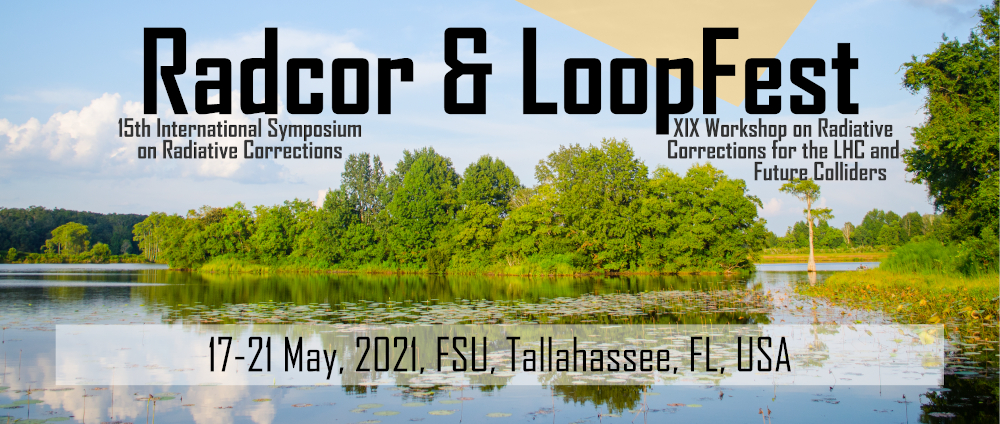}
  \end{minipage}
  &
  \begin{minipage}{0.85\textwidth}
    \begin{center}
    {\it 15th International Symposium on Radiative Corrections: \\Applications of Quantum Field Theory to Phenomenology,}\\
    {\it FSU, Tallahasse, FL, USA, 17-21 May 2021} \\
    \doi{10.21468/SciPostPhysProc.?}\\
    \end{center}
  \end{minipage}
\end{tabular}
}
\end{center}

\section*{Abstract}
{\bf
	We prove the conjecture made by Bern, Dixon, Dunbar, 
	and Kosower that describes 
	a simple dimension shifting relationship between the 
	one-loop structure 
	of $\mathcal{N}=4$ MHV amplitudes
	and all-plus helicity amplitudes in 
	pure Yang-Mills theory. 
	The proof captures all orders in dimensional
	regularisation using unitarity cuts, 
	by combining massive spinor-helicity with Coulomb-branch
	supersymmetry.
	The form of these amplitudes can be given in terms of pentagon 
	and box integrals using
	a generalised $D$-dimensional unitarity
	technique which captures the full amplitude 
	to all multiplicities.
}


\section{Introduction}
\label{sec:intro}
On-shell methods have lead to many new perspectives on gauge theories.
Not least of these is the gauge/gravity double copy~\cite{Bern2008,Bern2019}. 
Here we discuss and prove
an (up-until-now) conjectural relationship between gauge theories which relate a theory known to be physical, 
QCD,
to a theory of great theoretical and conceptual interest, 
$\mathcal N=4$ Super-Yang-Mills~\cite{Bern:1996ja}. 
This  states a simple dimension-shifting relationship
between the $n$-point one-loop gluon amplitudes of each theory $A^{\rm theory}_n$ with differing helicity configurations
\begin{align}
	A^{{\rm QCD}}_n\left(1^+,2^+,...,n^+\right) =
	-2\epsilon(1-\epsilon)(4\pi)^2\left[
	\frac{A_n^{\mathcal{N}=4}
	(1^+,...,i^-,...,j^-,...,n^+)}{
	\la ij \ra^4}\right]_{\epsilon\rightarrow \epsilon-2}
	\; ,
	\label{eq:conj}
\end{align}
where $\epsilon$ is the usual dimensional-regulator parameter and the $\la ij\ra^4$ factor
is the standard (Weyl) spinor-helicity contraction which compensates for the spinor-weight (or little-group scaling)
between the two amplitudes.

Of particular practical interest is the fact that the relation~\bref{eq:conj} holds to all orders in $\epsilon$, 
and thus relates the general-dimensional structure of the integral functions upon which it depends.
In~\cite{Bern:1996ja} the conjecture was verified up to $n=6$ but it has now been proven to all-multiplicities,
and to all orders in $\epsilon$~\cite{Britto2021}. Moreover, the complete all-orders-in-$\epsilon$ all-$n$ amplitudes 
can be computed in both theories.

\section{Status of the theories}
Since the conjecture of the relationship~\bref{eq:conj} 
was first made, 
there has been great progress in the computation 
of the relevant scattering amplitudes.
\subsection{All-plus QCD}
The all-plus QCD amplitude 
has long been known to vanish at tree-level order,
a fact which can most easily be seen using supersymmetric (SUSY) Ward identities~\cite{Grisaru:1977px}
\begin{align}
	A_n^{\rm SUSY;\;{tree}}(1^+,2^+,...,n^+) = 0 \quad ,
	\label{eq:apvan}
\end{align}
and the fact that at tree level
\begin{align}
	A_n^{\rm SUSY;\;{tree}}= A_n^{\rm QCD;\;{tree}} \; .
\end{align}

The one-loop order result was an early all-multiplicity result  for a gluon 
amplitude~\cite{Mahlon:1993si,Bern:1993qk}, 
and was computed to leading order in $\epsilon$
\begin{align}
	A_n^{\rm QCD,1-loop}(1^+,2^+,...,n^+)= \sum_{1\leq i_1<i_2<i_3<i_4\leq n}
	 {\tr_-(i_1i_2i_3i_4)\over \la 12...n1 \ra}
	 +\mathcal{O}(\epsilon)\quad .
	 \label{eq:qcd}
\end{align}
At two-loop order this particular helicity configuration provided the first high-multiplicity
($n>5$) results for the planar sector~\cite{Badger:2013gxa}, where 
thanks to four-dimensional cut-constructibility and on-shell recursion a functional form
has been successfully computed up to $n=7$ up to $\mathcal{O}(\epsilon )$ terms. 
A specific subleading-in-colour contribution to the two-loop all-plus amplitude
has recently been computed to two loops, the first partial-amplitude result  
to be computed to arbitrary multiplicity~\cite{Dunbar:2020wdh}.

\subsection{MHV in $\mathcal{N}=4$ SYM}
\begin{figure}[h]
\centering
\includegraphics{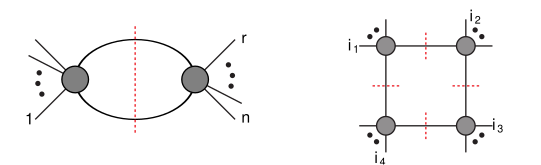}
	\caption[Cuts]{Unitarity cut constructions build loop amplitudes from tree amplitudes, and this 
	is particularly effective with the maximal supersymmetry of $\mathcal{N}=4$ SYM.}
    \label{fig:4dcuts}
\end{figure}
The tree-level amplitude of the MHV configuration in $\mathcal N=4$ SYM first deduced by Parke and 
Taylor~\cite{Parke1986} has since been expressed in a more general form, which bundles together 
the states in the supermultiplet~\cite{Bianchi:2008pu} into a "superamplitude"
\begin{align}
	A_n^{\rm MHV;\; tree}= {\delta^{(8)}\left(|i\ra \eta_{iA}\right)
\over \la 12\dots n1\ra} \; .
	\label{eq:delta}
	\intertext{Then the application of functional derivatives gives the tree amplitude with two negative-helicity gluons}
	A_n^{\mathcal{N}=4;\; {\rm tree}}(1^+,\dots,i^-,\dots,j^-,\dots,n^+)
	&={\delta^4\over \delta \eta_i^4}
	{\delta^4\over \delta \eta_j^4}A_n^{\rm MHV;\;  tree} \; .
\end{align}

At loop-level, the MHV amplitude is one of the most resounding successes of four-dimensional unitarity constructions. 
The amplitude is easily constrained and fixed by unitarity diagrams like those on the left-hand side of Figure~\ref{fig:4dcuts}.
Together with the simple form of the tree amplitude, this resulted in 
another amplitude to have been computed up to $\mathcal{O}(\epsilon)$ to arbitrary multiplicity~\cite{Bern:1994zx,Bern:1994cg}:
\begin{align}
	A_n^{\rm MHV} = {1\over 4}{\delta^{(8)}
	(|i\ra \eta_{iA})
	\over \la 12...n1\ra}
	\sum_{i_1,i_3=1}^n
	\tr(i_1q_{i_1+1,i_3}i_3q_{i_3+1,i_1})
	I_4^{[i_1,i_1+1,i_3,i_3+1]}
	+\mathcal{O}(\epsilon)\; ,
	\label{eq:n4trunc}
\end{align}
where $q_{ij}=p_i+\dots+p_{j-1}$ where the counting is defined cyclically in terms of particle labels.
The amplitude~\bref{eq:n4trunc} can also very easily be computed using generalised unitarity cuts~\cite{Britto:2004nc} which builds
the entire amplitude from on-shell products of amplitudes like 
the one depicted on the right-hand-side of Figure~\ref{fig:4dcuts}.

Astoundingly, multiloop results for $(n<6)$-points extend to all orders, thanks to the exponentiation of the one-loop result given by 
the Bern-Dixon-Smirnov (BDS) ansatz~\cite{Bern2005}.
For higher $n$ the problem then turns into fixing the rest of 
the structure not captured by the BDS ansatz. 
Two-loop results have recently 
reached $n=9$~\cite{Golden2021} 
and $n=6$ results extending through to seven
loops~\cite{Caron-Huot:2019bsq}. 

\section{Proving the conjecture}
The verification of the relationship~\bref{eq:conj} was done 
in~\cite{Bern:1996ja} to all
orders up to $n=6$ multiplicity. This combined a string-derived formalism for the $\mathcal N=4$ side~\cite{Bern:1991aq},
and $D$-dimensional unitarity for the QCD side. 
We use the $D$-dimensional unitarity on both sides, matching the cuts which capture the full amplitude
to all orders in $\epsilon$.
 
\subsection{Statement in terms of $D$-dimensional cuts}
To compute full amplitudes in dimensional regularisation we can use a technique equivalent to taking 
massive unitarity cuts.
$D$-dimensional unitarity treats a cut in $4-2\epsilon$ dimensions by splitting up the loop momentum
into a four-dimensional contribution
$l$ (which lives in the same space as the external momenta) and the $-2\epsilon$ difference
\begin{align}
	\ell = l + \ell^{[\mi 2\epsilon]}\; ,
\end{align}
and the unitarity-cut (on-shell) condition becomes
\begin{align}
	\ell^2 = l^2 - \mu^2 = 0 \; ,
\end{align}
where we have defined 
\begin{align}
\mu^2 \equiv \left(\ell^{[\mi 2\epsilon]}\right)^2 \;.
	\label{eq:mass}
\end{align}
Meanwhile, as was originally observed in~\cite{Bern:1996ja},
the vanishing of all-plus gluon amplitudes in theories with supersymmetry 
implies that 
\begin{align}
	A_n^{\rm QCD}(1^+,2^+,\dots,n^+)= 2A_n^{[0]}(1^+,2^+,\dots,n^+)
\end{align}
where $A_n^{[0]}$ is the amplitude with real scalar gluons circulating in the loop.
This implies that the unitarity cuts are given by the product of
scalar-gluon tree amplitudes, where the scalars have mass as 
defined in equation~\bref{eq:mass}.

In particular, the dimension-shifting relationship can be stated 
at the integrand level in terms of the parameter $\mu^2$.
In considering the full integral in dimensional regularisation
we see that the "dimension shift" comes from 
$\mu^2$ terms present in the numerator
\begin{align}
	I^{4-2\epsilon}\left[\mu^{2r}\right] &=
	 -\int{d^4ld\mu^2 \over (-\mu^2)^{1+\epsilon}}
	 {\left(\mu^2\right)^r\over \left(l^2-\mu^2\right)
	 \cdots\left((l-q)^2-\mu^2\right)}
	 \notag \\
	&=
	-\epsilon(1-\epsilon)\cdots (r-1-\epsilon)
	I^{4+2r-2\epsilon}\left[1\right] 
\end{align}
which means we can restate~\bref{eq:conj} in terms of a 
relationship between unitarity cuts.
Thus, as suggested in~\cite{Bern:1996ja} we prove the relationship
\begin{align}
	A^{\rm QCD}_n\biggr|^{\mu^2\neq 0}_{q_{rs}\; {\rm cut}} 
	= A^{\mathcal{N}=4}
	\left[{2\mu^4\over \la ij\ra^4}\right]\biggr|^{\mu^2\neq 0}_{q_{rs}\; {\rm cut}}
	\; 
	\label{eq:cutconj}
\end{align}
for cuts in all momentum channels $q_{rs}^2$.
\subsection{Necessary tree amplitudes}
Proving equation~\bref{eq:cutconj} requires 
understanding the $D$-dimensional cuts on the 
$\mathcal{N}=4$ side. 
The amplitudes needed are 
known~\cite{Forde:2005ue,Rodrigo:2005eu,Ferrario:2006np,Kiermaier:2011cr}, and 
through Coulomb-branch supersymmetry they can 
be bundled together
into the "MHV-band" amplitudes~\cite{Craig:2011ws,Kiermaier:2011cr,Elvang:2011ub}, which admit 
a delta-function representation analogous to the one given
in equation~\bref{eq:delta},
\begin{align}
	A_{\rm tree}^{\rm MHV-band} = 
	{[\lambda_n\lambda_1]^2	\delta^\chi_{12}\delta^\chi_{34}
				\over  m^2q_{n2}^4}
	A(\mathbf{1}^0,2^+,3^+,
			...,(n-1)^+,\mathbf{n}^0) \; ,
		\label{eq:mhvband}
\end{align}
where $A(\mathbf{1}^0,2^+,3^+,...,(n-1)^+,\mathbf{n}^0)$
are precisely the massive-scalar-gluon amplitudes needed for the 
QCD side of equation~\bref{eq:cutconj}. Here $q_{n2}=p_n+p_1$,
and $[\lambda_i|$ are the spinors obtained from the projection of $p_i$ against
a null reference spinor (or $q$-frame) $p_\chi$ in the Dittmaier massive-spinor-helicity formalism~\cite{Dittmaier:1998nn}.
The delta functions $\delta_{ij}^\chi$ are defined in~\cite{Craig:2011ws}. They depend on the reference momentum $p_\chi$ 
and encode the remaining supersymmetry after it is partially broken in the Coulomb branch by the 
introduction of the mass $m$~\cite{Craig:2011ws}.

\subsection{Proof}
\begin{figure}[h]
\centering
\includegraphics{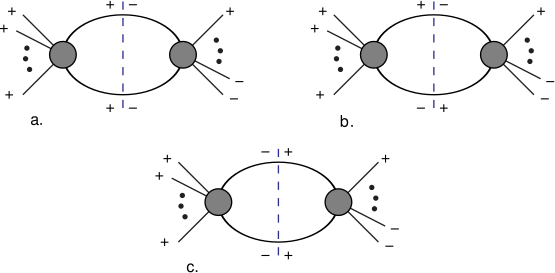}
	\caption[proog]{The three types of cut which reproduce the $\mathcal N=4$ amplitude.}
    \label{fig:cuts}
\end{figure}
The tree amplitudes which can be extracted from equation~\bref{eq:mhvband} are most compactly expressed in
the gauge constrained such that $p_{\chi}\cdot q_{n2}=0$. We consider the sum of the helicity states in this gauge
which correspond to the configurations of the general types depicted in Figure~\ref{fig:cuts}.

 Cuts of type (a) vanish as a consequence of the vanishing of the all-plus-helicity tree amplitudes on the left-hand side. 
Cuts of type (b) give a non-vanishing contribution for the fermionic, gluonic,
and scalar loop content in the supermultiplet, but the total cancels so that these cuts do not contribute to 
the full amplitude. 

So the only type of cuts contributing are (c), which are MHV on both sides and capture 
the epsilon-truncated one-loop amplitude in equation~\bref{eq:n4trunc}. In this case they are actually capturing 
the structure to all orders in $\epsilon$ through the Coulomb-branch amplitudes.

The proof then proceeds as follows. The cut integral is
\begin{align}
	\int &d^4\eta_{l_r}d^4\eta_{l_s}
		A^{\rm MHV;\; tree}_L(\mi{\bm l}_r^1,r,
		...,(s-1),{\bm l}_{s}^1)
		\times
	A^{\rm MHV;\; tree}_R(\mi{\bm l}_r^1,r,...,s-1,{\bm l}_s^1)
	=
	\notag \\
	&\int d^4\eta_{l_r}d^4\eta_{l_s}
		{\delta^{(8)}\left(L\right)\over \mu^2 \la 
		\lambda_{l_s}\lambda_{l_r}\ra^2}
		A_L(\mi{\bm l}_r^0,r,
		...,(s-1),{\bm l}_{s}^0)
		\times
	{\delta^{(8)}\left(R\right)\over \mu^2 \la 
		\lambda_{l_s}\lambda_{l_r}\ra^2}
	A_R(\mi{\bm l}_r^0,s,...,r-1,{\bm l}_r^0) \; ,
	\notag \\
	&\hspace{0.5in}L\equiv|i \ra \eta_{iA}, \;i\in \lbrace \lambda_{\mi l_r},
		r,...,	s-1,\lambda_{l_s}\rbrace
		\; ; \;
	R\equiv	|i	 \ra 
		\eta_{jA} ,\; i\in \lbrace \lambda_{l_s},
		r,...,
		s-1,\lambda_{\mi l_r}\rbrace\; ,
\end{align}
and applying the Grassmann integration and delta functions gives
\begin{align}
	A^{\rm MHV}\biggr|_{q_{rs}\;{\rm cut}}=&
	{\delta^{(8)}\left( \la \lambda_i | \eta_{iA}\right)
	\over \mu^4}A_L(\mi{\bm l}_r^0,r^+,...,(s-1)^+,{\bm l}_s^0)
	A_R(\mi{\bm l}_s^1,s^+,...,(r-1)^+,{\bm l}_r^1) \; .
\end{align}
Applying the 
functional derivatives then gives
${\delta^4\over\delta\eta_i^4} 
	{\delta^4\over\delta\eta_j^4}$, 
\begin{align}
	A_n^{\mathcal{N}=4}(1^+,2^+,...,i^-,...,j^-,...,n^+)
	\biggr|_{q_{rs}\;{\rm cut}} 
	&={\delta^4\over\delta\eta_i^4} 
	{\delta^4\over\delta\eta_j^4}\left[ A^{\rm MHV}_n
	\biggr|_{q_{rs}\;{\rm cut}}  
	\right]
\notag \\
	&= {\la ij \ra ^4\over 2\mu^4}A_n^{\rm QCD}
	\biggr|_{q_{rs}\;{\rm cut}} 
\end{align}
which proves the conjecture~\bref{eq:conj}.

\section{Closed forms to all multiplicities}
Through the conjecture, we need only compute one side to get the general all-orders-in-$\epsilon$, all-$n$ form of the amplitude
in both theories. This amounts to extracting coefficients of box and pentagon integrals. 

\subsection{Box coefficients}
These coefficients are remarkably simple to compute on the $\mathcal N=4$ side, 
as the boxes are given from four-dimensional cuts and  we can read the coefficients
directly from equation~\bref{eq:n4trunc}. 
Namely a coefficient $b_4$ is given by 
\begin{align}
	b_4^{{\rm QCD};\;[i_1,i_3 -  1,i_3,i_1 -  1]} &= 
	 {1\over 2} {\tr(i_1q_{i_1i_3}i_3q_{i_3+1,i_1-1})\over
	\la 12 ...n1\ra }\; ,
	\label{eq:b4}
\end{align}
but we can also see these emerge on the QCD side from imposing ultraviolet constraints on the box-integral
basis and demonstrating that only "two-mass-easy" ($i_2=i_3-1$, 
$i_4=i_1-1$) box coefficients contribute. 
The coefficients are then given by the formula 
\begin{align}
	b_4^{{\rm QCD};\; [i_1,i_3 -  1,i_3,i_1 -  1]} &= 
	{1\over \mu^4} \left[2\times {1\over 2} \sum_{\alpha_\pm}
	A^{\rm tree}
	\times 
	A^{\rm tree}
\times
	A^{\rm tree}
	\times A^{\rm tree}
	\biggr|_{\mathcal{O}(\mu^6)}\right] .
\end{align}
\subsection{Pentagon coefficients}
\begin{figure}[h]
\centering
\includegraphics{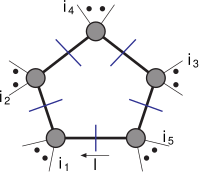}
	\caption[penta]{The five-particle cut reproduces pentagon coefficients.}
    \label{fig:pentacut}
\end{figure}
The pentagon coefficients can be easily solved for thanks to the generalised $D$-dimensional unitarity penta-cut solution
presented in~\cite{Jehu:2020xip}, 
which simply gives an explicit solution for the loop momentum given five massive (or $D$-dimensional)
cuts depicted in Figure~\ref{fig:pentacut}
\begin{align}
	l_{i_1}^\mu&=-{\tr_5\left(q_{i_1i_2}q_{i_2i_3}
	q_{i_3i_4}q_{i_4i_5}q_{i_5i_1}\gamma_\mu\right)\over 
			2\tr_5(q_{i_1i_2}q_{i_2i_3}
	q_{i_3i_4}q_{i_4i_5})}
	\notag \\
			\mu^2 &= 
			{\tr(q_{i_1i_2}q_{i_2i_3}
	q_{i_3i_4}q_{i_4i_5}q_{i_5i_1}q_{i_1i_2}q_{i_2i_3}
	q_{i_3i_4}q_{i_4i_5}q_{i_5i_1})- 2\prod_{k=1}^5
	q_{i_ki_{k+1}}^2\over
				\tr_5^2(q_{i_1i_2}q_{i_2i_3}
	q_{i_3i_4}q_{i_4i_5})}
						\label{eq:5cutsol}
\end{align}
and this need only be plugged into the product of the five on-shell amplitudes to give the coefficients:
\begin{align}
	b_5^{{\rm QCD};\; [i_1,i_2,i_3,i_4,i_5]} = c_0^3 
	A^{\rm tree}(\mi{\bm l}^0_{i_1}, i_1^+,...,(i_2- 1)^+,
	{\bm l}^0_{i_2})\times
	A^{\rm tree}(\mi{\bm l}^0_{i_2},i_2^+,...,(i_3- 1)^+,
	{\bm l}^0_{i_3})&\times
	\notag \\
	A^{\rm tree}(\mi{\bm l}^0_{i_3},i_3^+,...,(i_4- 1)^+,{\bm l}^0_{i_3})\times 
	A^{\rm tree}(\mi{\bm l}^0_{i_4},i_4^+,...,(i_5- 1)^+,{\bm l}^0_{i_3})&\times
	\notag \\
	A^{\rm tree}(\mi{\bm l}^0_{i_5},i_5^+,...,(i_1- 1)^+,{\bm l}^0_{i_3})&\quad .
	\label{eq:b5}
\end{align}
A non-trivial check is that this reproduces the parity-odd contributions to the finite QCD amplitude, 
namely the $\tr_5$ piece of the QCD amplitude in equation~\bref{eq:qcd} where
\begin{align}
	\tr_-(i_1i_2i_3i_4) = {1\over 2}\left(\tr(i_1i_2i_3i_4)
	-\tr_5(i_1i_2i_3i_4)\right)
\end{align}
matches the $b_5$ contributions upon substituting the pentagon integrals for their values in the $\epsilon \to 0$ limit 
\begin{align}
	I_5[\mu^4]\underset{\epsilon = 0}{\rightarrow}-{1\over 24} \; . 
\end{align}
We confirm this numerically up to to the $n=17$ case.

\section{Conclusion}
The dimension shift relationship between QCD all-plus-helicity amplitudes and 
$\mathcal N=4$ SYM MHV amplitudes has been proven at one-loop to all multiplicities. 
We have also given all-$n$ all-orders-in-$\epsilon$ expressions in terms of
pentagon and box integrals, through fixing their coefficients through generalised $D$-dimensional unitarity cuts.
The origin of the box coefficients is particularly distinct, 
with the $\mathcal N=4$ computation falling out automatically from four-dimensional
cuts while the QCD requires UV truncation of the integral basis in $D$ dimensions to fully constrain.

Further work towards studying the analytic structure of these amplitudes to all orders in $\epsilon$ 
could yield deeper insight at higher-loop order. In particular, we observe that the "purity" (polylogarithmic
simplicity) of the amplitude is broken at high orders in $\epsilon$ in $\mathcal N=4$ super Yang-Mills. 
Future work will involve getting a stronger analytic control of these all-$n$ all-orders-in-$\epsilon$ expressions.

\section*{Acknowledgements}
Diagrams were drawn using Axodraw.
This project has received funding from the European Research Council (ERC) under 
the European Union’s Horizon 2020 research and innovation programme (Grant agreement No. 647356 “CutLoops.") 




\bibliographystyle{SciPost_bibstyle} 
\bibliography{biblo.bib}

\nolinenumbers

\end{document}